\documentclass[aps,prl,preprint,groupedaddress,superscriptaddress]{revtex4-1}
\usepackage{graphicx, subfigure}
\usepackage[pdfborder={0 0 0.5}]{hyperref}
\usepackage{epstopdf}
\usepackage{amsmath}
\usepackage{siunitx}

\begin{document}

\preprint{}

\title{Ferroelectricity and Rashba effect in GeTe}


\author{C. Rinaldi}
\email{christian.rinaldi@polimi.it}
\affiliation{Dipartimento di Fisica, Politecnico di Milano, Via Colombo 81, 20133 Milano, Italy}
\author{D. Di Sante}
\affiliation{Consiglio Nazionale delle Ricerche CNR-SPIN, UOS L'Aquila, Via Vetoio 10, 67100 L'Aquila, Italy}
\affiliation{Department of Physical and Chemical Sciences, University of L'Aquila, Via Vetoio 10 67100, L'Aquila, Italy}

\author{A. Giussani}
\author{R.-N. Wang}
\affiliation{Paul-Drude-Institut f{\"u}r Festk{\"o}rperelektronik, Hausvogteiplatz 5-7, 10117 Berlin, Germany }

\author{S. Bertoli}
\author{M. Cantoni}
\author{L. Baldrati}
\affiliation{Dipartimento di Fisica, Politecnico di Milano, Via Colombo 81, 20133 Milano, Italy}

\author{I. Vobornik}
\author{G. Panaccione}
\affiliation{Consiglio Nazionale delle Ricerche, CNR - IOM, Laboratorio TASC, I-34149 Trieste, Italy}

\author{R. Calarco}
\affiliation{Paul-Drude-Institut f{\"u}r Festk{\"o}rperelektronik, Hausvogteiplatz 5-7, 10117 Berlin, Germany }

\author{S. Picozzi}
\affiliation{Consiglio Nazionale delle Ricerche CNR-SPIN, UOS L'Aquila, Via Vetoio 10, 67100 L'Aquila, Italy}
\author{R. Bertacco}
\affiliation{Dipartimento di Fisica, Politecnico di Milano, Via Colombo 81, 20133 Milano, Italy}


\date{\today}

\begin{abstract}
GeTe has been proposed as the father compound of a new class of functional materials displaying bulk Rashba effects coupled to ferroelectricity: ferroelectric Rashba semiconductors. In nice agreement with first principle calculations, we show by angular resolved photoemission and piezo-force microscopy that GeTe displays surface and bulk Rashba bands arising from the intrinsic inversion symmetry breaking provided by the remanent ferroelectric polarization. This work points to the possibility to control the spin chirality of bands in GeTe by acting on its ferroelectric polarization.
\end{abstract}


\maketitle
 Spinorbitronics, i.e. the use of spin-orbit coupling as key ingredient in electronic devices, is an emerging route in spintronics \cite{Manchon2014}. After the first demonstration of new concepts, such as spin-orbit torque for switching the magnetization via momentum transfer from a current flowing in a high spin-orbit material \cite{Garello2013}, it is nowadays clear that new materials are needed in order to bridge the gap towards applications. Besides, using spin-orbit effects in semiconductors represents an appealing strategy to implement a spin transistor, the spin-based analogue of the transistor which represents the "holy grail" for spintronics. However, even in the case of the well-known "Datta-Das" spin-FET architecture \cite{Datta1990}, so far there have been only few works dealing with the feasibility of this approach in two-dimensional electron gas structures \cite{Koo2009}. Therefore, semiconductors showing tunable bulk spin-orbit related transport effects still represent a high ambition in spintronics.
\\
\indent In this paper we investigate the intriguing properties of GeTe, the father compound of a new class of multifunctional materials: ferroelectric Rashba semiconductors (FERSCs). They are predicted to display an intrinsic link between ferroelectric (FE) polarization and spin chirality in Rashba bands, thus paving the way to the electric control of spin transport properties \cite{Picozzi2014,DiSante2013}. By piezo-force microscopy (PFM) and angular resolved photoemission spectroscopy (ARPES) we show that rhombohedrally distorted $\alpha$-GeTe(111) thin films present a macroscopic built-in FE polarization leading to Rashba splitting of surface and bulk bands, in excellent agreement with calculations based on density functional theory (DFT). 
\begin{figure}
	\includegraphics[width=8.4cm]{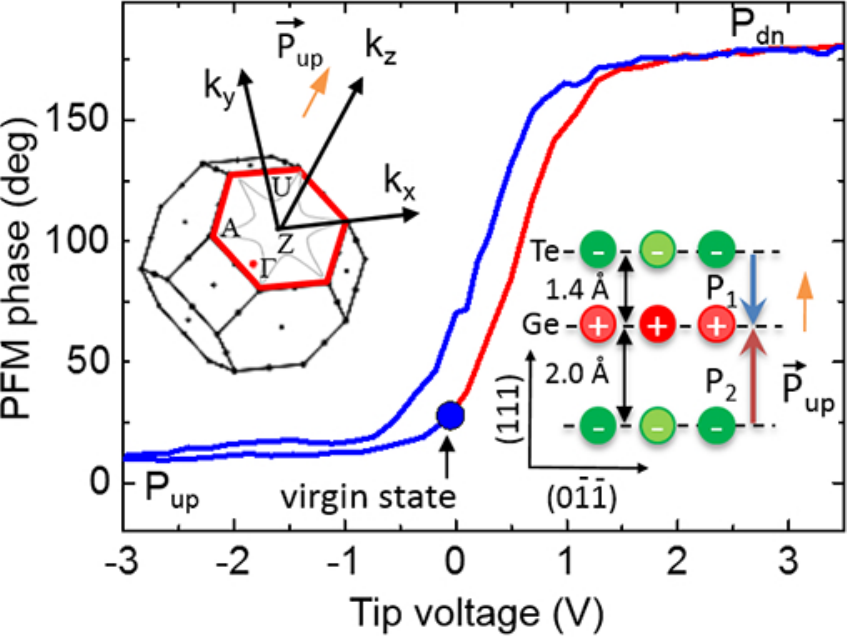}%
	\caption{\label{fig:Fig1} PFM hysteresis loop measured by sweeping the tip voltage (zero, positive, negative, zero) from the virgin state. The upper left inset shows the BZ of GeTe. The grey six-fold star indicates the symmetry of isoenergy cuts of bulk Rashba bands at 0.3 eV below the VBM from Ref. \cite{DiSante2013}. The bottom right inset shows that short and long bonds between Te and Ge planes result in a net outward FE polarization of the GeTe(111) surface.}%
\end{figure}

Below 720 K, GeTe displays a non-centrosymmetric ferroelectric structure (space group R3m (No. 160)), with Ge and Te ions displaced from the ideal rocksalt sites along the [111] direction ,so as to display a remanent FE polarization \cite{Chattopadhyay1987,Rabe1987,Shaltaf2008}. In Figure \ref{fig:Fig1}, a sketch of the GeTe bulk Brillouin zone (BZ) in the rhombohedral setting is reported. According to Ref. \cite{DiSante2013}, a giant Rashba dispersion is expected in the plane of the hexagonal face (red hexagon in Fig. \ref{fig:Fig1}) centered around the Z point and perpendicular to the [111] direction ($k_z$), especially along the high symmetry directions ZA ($k_x$) and ZU ($k_y$). To expose the correct surface to ARPES investigation, we used $\alpha$-GeTe(111) $\SI{20}{\nano\meter}$ thick films, grown by molecular beam epitaxy (MBE) \cite{Giussani2012} on Si(111) slightly p-doped substrates. After growth, GeTe films have been capped with Si$_3$N$_4$ in the MBE chamber to avoid contamination upon exposure to atmosphere. Si$_3$N$_4$ stripping has been performed by wet etching in hydrofluoric acid, before the introduction in the APE beamline station of the Elettra synchrotron radiation facility. Then the sample has been annealed up to  $\SI{250}{\degreeCelsius}$ in UHV to promote contaminants desorption and surface ordering.

The first step in the investigation of the FERSC features is the FE characterization of GeTe thin films. Despite GeTe represents one of the simplest cases of displacive FE, so far there are only a few papers reporting on its FE behavior \cite{Pawley1966,Matsunaga2008,Kadlec2011,Gervacio-Arciniega2012,Polking2012}. Recently Kolobov \emph{et al.} \cite{Kolobov2014} have shown by PFM that $\alpha$-GeTe(111) epitaxial films, like those used in this paper, but still capped with Si$_3$N$_4$, display FE remanence. We used an Agilent 5600 microscope operated in PFM mode to probe the FE properties of GeTe films which underwent the very same surface preparation used for ARPES. In this way we can correlate the structure of Rashba bands measured by ARPES and the FE polarization. PFM point spectroscopy was initially performed to check the virgin state polarization, as illustrated in Fig. \ref{fig:Fig1}. A clear hysteresis loop in the phase signal is present, with the virgin state polarization having the same sign and similar magnitude than that written with negative bias applied to the AFM tip. This indicates an intrinsic tendency of the films to display an outward polarization ($P_{up}$). The remanence was quite low in these thin films after Si$_3$N$_4$ removal, but still large enough to write and detect FE domains, thus confirming the existence of an outward polarization over large areas. This arises from the thermodynamic stability of the Te-terminated surface of $\alpha$-GeTe(111), as resulting from DFT calculations \cite{Deringer2012}, which promotes this termination (see inset of Fig. \ref{fig:Fig1}) \cite{Wang0}. As a result of the shorter Ge-Te bonds' arrangement at the very surface, the intrinsic surface FE dipole tends to point outwards ($P_{up}$ in the inset of Fig. \ref{fig:Fig1}), as experimentally observed by PFM. This provides the macroscopic symmetry-breaking necessary to observe the k-splitting of surface Rashba bands in photoemission.

\begin{figure}
	\includegraphics[width=8.6cm]{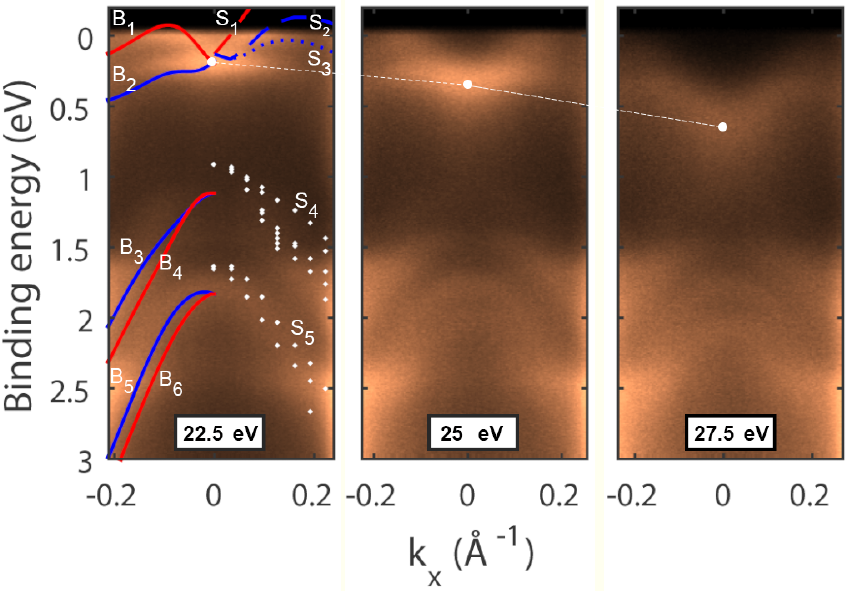}%
	\caption{\label{fig:Fig2} Band dispersions along $k_x$ ($ZA$,$\overline{\Gamma} \overline{K}$) for 22.5, 25 and 27.5 eV photon energy along with bulk Rashba bands (B$_1$ and B$_2$) from Ref. \cite{DiSante2013} and surface Rashba bands (S$_1$, S$_2$, S$_3$, S$_4$) as calculated for a relaxed Te-terminated GeTe surface (see also Fig. \ref{fig:Fig4}).}
\end{figure}

In order to investigate the dispersion of Rashba bands by ARPES, we first selected the right photon energy at which photoelectrons have initial $k_z$ close to the Z point in normal emission. As the VBM in Z corresponds to the absolute VBM of $\alpha$-GeTe,\cite{DiSante2013} the correct value of $h\nu$ is that for the topmost occupied band seen in photoemission is closer to the Fermi level. To this scope we measured $E\left( k_x \right)$ bands dispersions at various photon energies ($h\nu$), from 22.5 to 40 eV, in such a way to cover half of the BZ along Z$\Gamma$. In Fig. \ref{fig:Fig2} we concentrate on dispersions taken at 22.5, 25.0 and 27.5 eV photon energy. For a deeper understanding, we compare experimental dispersions with theoretical Rashba bands. A general nice agreement is found with bands calculated by DFT in the rhombohedral setting, as visible in Fig. \ref{fig:Fig2}a, which reports the superposition of bulk (continuous line for negative momenta) and surface (dashed lines for positive momenta) bands. Red and blue colors indicate bands with spin perpendicular to the wave vector and circulating clockwise or counter clockwise around the [111] direction. Close to $E_F$ two bands are seen, displaying the typical Rashba zero k-crossing as well as a downward dispersion as a function of $h\nu$ which indicates their bulk character. They correspond to bulk Rashba bands of Ref. \cite{DiSante2013} (B$_1$ and B$_2$ from now on). Note that part of the signal seen at 22.5 eV around $E_F$ comes from surface Rashba bands $S_1$, $S_2$ and $S_3$, not dispersing with $h\nu$. However, the minimum energy distance from $E_F$ of B$_1$ and B$_2$ is found at about 22.5 eV: this is the photon energy we used to probe states around the $Z$ point.

\begin{figure*}
	\includegraphics[width=17.2cm]{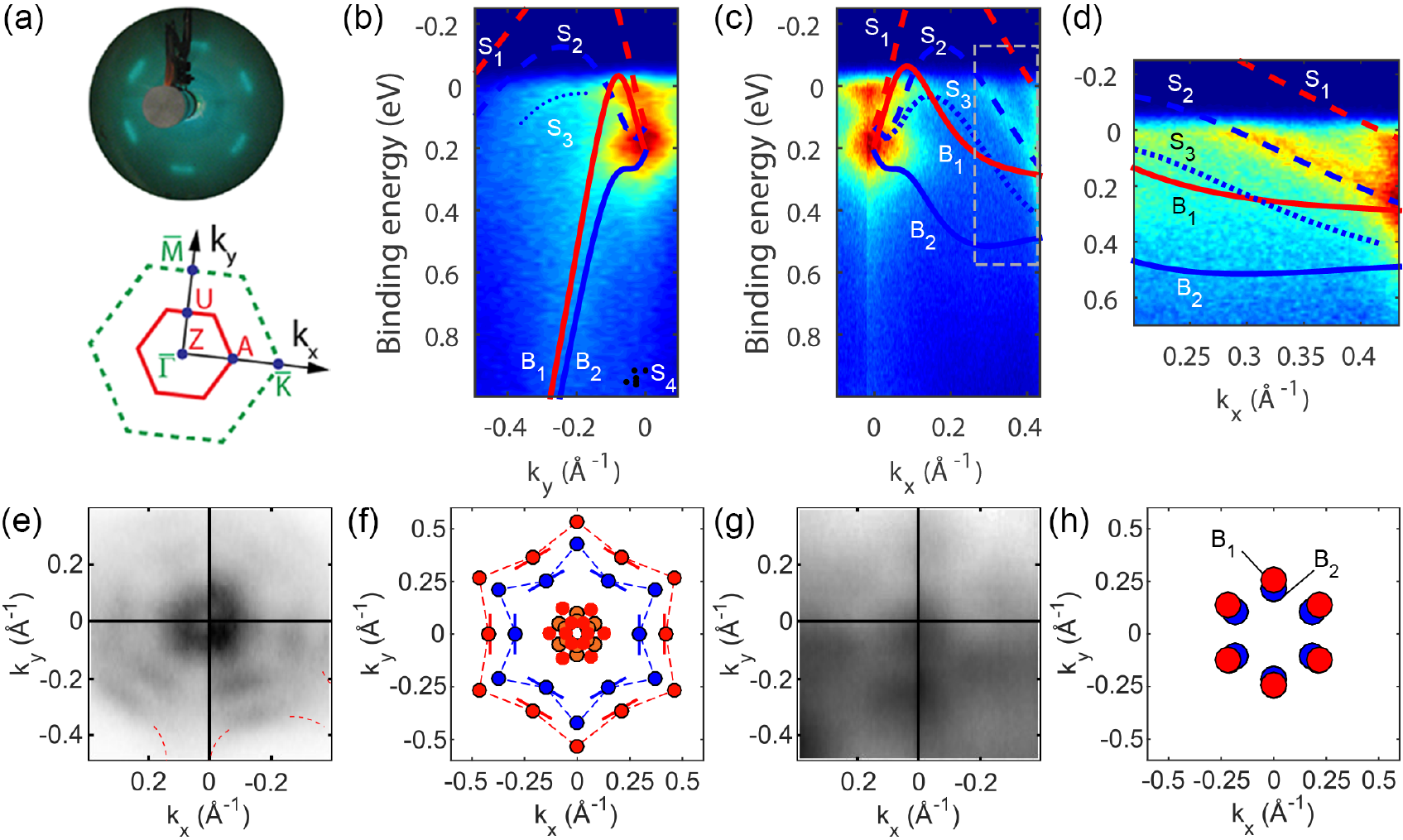}
	\caption{\label{fig:Fig3} (a) Top: LEED taken at 80 eV; bottom: surface BZ (green) and hexagonal face of the 3D BZ (red) (b) Band dispersion along $k_y$ ($ZU$,$\overline{\Gamma} \overline{M}$). (c) Band dispersion along $k_x$ ($ZA$,$\overline{\Gamma} \overline{K}$). (d) Zoom of panel 3c in the dashed line rectangle. (e) Experimental isoenergy cut at $E_F$. (f) Simulated isoenergy cut at $E_F$ from theoretical Rashba bands. Parallel segments indicate the angular broadening of crossing points along ($ZA$,$\overline{\Gamma} \overline{K}$) due to rotational domains. In the inner part, the intersection of the B$_1$ and B$_2$ Rashba bands along $ZU$ and $ZA$ are shown. (g) Experimental isoenergy cut at 0.8 eV BE. (h) Simulated isoenergy cut from theoretical bulk Rashba bands at 0.8 eV BE.}
\end{figure*}

Note that the band assignment and the corresponding fit of experimental data with DFT bands of Fig. \ref{fig:Fig2} corresponds to a VBM crossing $E_F$. This is not surprising due to the high concentration of Ge vacancies in single crystal GeTe (about 10\%), which act as acceptors. Together with local distortions they are essential for the stability of the crystalline phase as shown by Wuttig \emph{et. al.} \cite{Wuttig2007}. Indeed, Hall measurements on very similar GeTe films indicate a heavy bulk p-doping with hole concentration of $\SI{5.3e20}{\per\cubic\centi\meter}$.

Fig. \ref{fig:Fig3}a shows a low-energy electron diffraction (LEED) pattern taken at 80 eV from the sample surface prepared at the APE station. A good pattern with the expected sixfold symmetry is found after surface preparation in vacuum but with about $\SI{14}{\degree}$ angular broadening around $k_z$, arising from the presence of five distinct rotational domains.\cite{Giussani2012} The surface BZ (dashed green hexagon) and the projection of the hexagonal face of the bulk BZ (red hexagon) are also shown at the bottom of Fig. \ref{fig:Fig3}a, where both bulk and surface high symmetry points are reported. 

\begin{figure}
	\includegraphics[width=8.6cm]{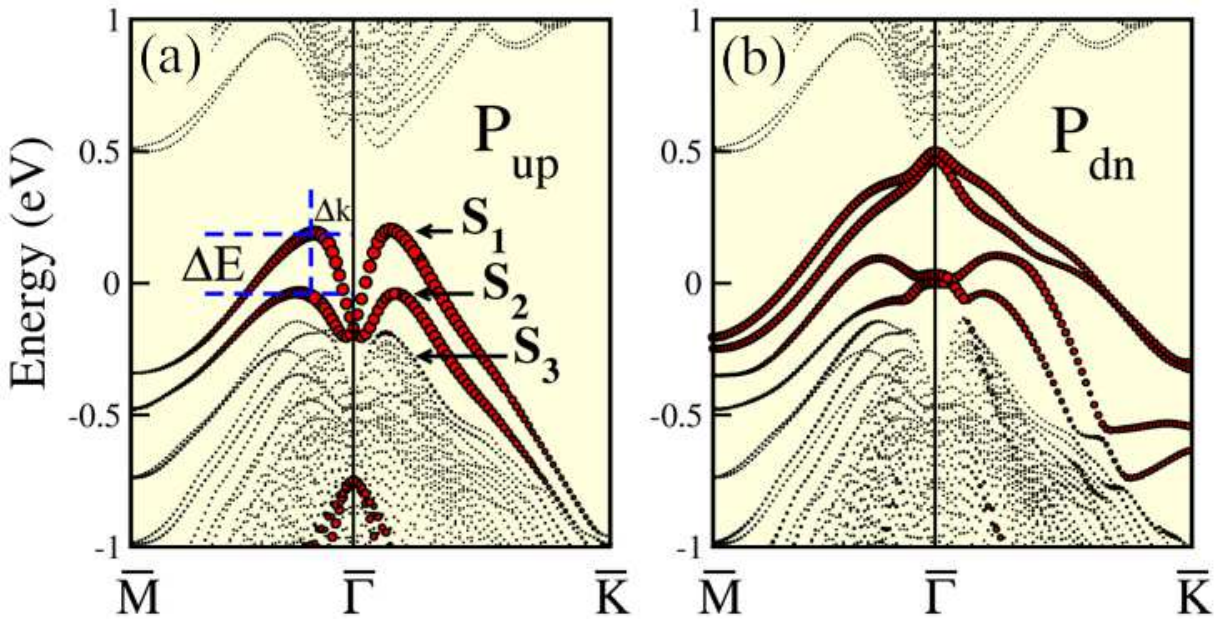}%
	\caption{\label{fig:Fig4} Theoretical surface Rashba bands calculated for a Te terminated surface with the polarization $P_{up}$ (panel a) and $P_{dn}$ (panel b). Red circles indicate bands with a relevant weight (proportional to the circle radius) arising from surface Te atoms. Small black dots highlight the continuum of bulk states in the supercell calculation.} %
\end{figure}

In Fig. \ref{fig:Fig3}b and \ref{fig:Fig3}c we report the band dispersion along $k_y$ ($ZU$,$\overline{\Gamma}\overline{M}$) and $k_x$ ($ZA$,$\overline{\Gamma} \overline{K}$) down to 1 eV binding energy, measured at 4 K, as compared with bulk and surface Rashba bands. Along $k_y$ (Fig. \ref{fig:Fig3}b) the steeply lateral falling signal can be attributed to B$_1$ and B$_2$ bulk Rashba bands, while surface states S$_1$ and S$_2$ cross $E_F$ at low momenta (lower than $\SI{0.12}{\per\angstrom}$) and mainly develop in unoccupied states. A tiny signal from a third surface band (S$_3$) is also seen (dotted line), almost tangent to the Fermi level. Note that the spectral weight of this band predicted by DFT is much lower than that of S$_1$ and S$_2$ (see Fig. \ref{fig:Fig4}) and this results in a lower intensity, as it is clearly observed in Fig. \ref{fig:Fig3}c.  Along $k_x$ (Fig. \ref{fig:Fig3}c), there is only a diffuse signal arising from bulk (B$_1$ and B$_2$) bands, while the two parallel surface Rashba bands S$_1$ and S$_2$, re-entering from unoccupied states above $\SI{0.2}{\per\angstrom}$, are clearly detected. This is more evident in the zoom of Fig. \ref{fig:Fig3}c reported in panel 3d, where also the presence of some signal from the lowest surface band S$_3$ and the bulk band B$_1$ and B$_2$ is seen.

The Rashba character of surface bands S$_1$ and S$_2$ can be better appreciated from the isoenergy cut of Fig. \ref{fig:Fig3}e, taken at the Fermi level ($E_F$). The six couples of parallel segments centred along the (ZA,$\overline{\Gamma}\overline{K}$) directions arise from the crossing of the Fermi surface by S$_1$ and S$_2$ which re-enter from unoccupied states and provide proof of the Rashba character of such bands. Despite their angular broadening, resulting from rotational domains visible in the LEED pattern shown in Fig. \ref{fig:Fig3}a, a clear k-splitting is detected. In particular they can be associated to the central portions between the "arms" of the six-fold stars which are found by cutting the theoretical surface S$_1$ and S$_2$ Rashba bands at $E_F$. This is shown in panel \ref{fig:Fig3}f, where a distance of the segments from $\overline{\Gamma}$ of 0.3 and $\SI{0.4}{\per\angstrom}$ is found, in nice agreement with experimental cuts. The "arms" of the theoretical surface star at $E_F$, instead, stretch along (ZU,$\overline{\Gamma} \overline{M}$) but their intersection with the Fermi surface is located at higher momenta, so that we can only see some signal due to these arms at negative $k_y$ (see dashed lines in panel \ref{fig:Fig3}e). 
Note also the central hole in the isoenergy cut of Fig. \ref{fig:Fig3}e, which is an additional indication of Rashba bands with the characteristic k-crossing at zero momentum. The signal around this central hole still has the shape of a sixfold star, but this time the arms stretch along ZA (see the grey star in the left-top inset of Fig. \ref{fig:Fig1}a), as expected for bulk Rashba bands. As a matter of fact, when looking at both panels \ref{fig:Fig3}b and \ref{fig:Fig3}c, we clearly see that the outermost band crossing the Fermi surface is B$_1$, along $k_x$ (ZA). This is visible also in the inner part of panel \ref{fig:Fig3}f, showing the expected intersection of the theoretical bulk bands B$_1$ and B$_2$ with the Fermi surface. The orientation of the inner star of Fig. \ref{fig:Fig3}e is thus determined by bulk states. This is rotated by 30 degrees with respect to the outer star associated to surface states. 

To probe only bulk Rashba bands in Fig. \ref{fig:Fig3}g we plot the isoenergy cut at 0.8 eV BE, where surface states are not expected (see Fig. \ref{fig:Fig3}b-c). A central circle can be distinguished, arising from the cut of the top of surface band S$_4$, as well as six circles at about $\SI{0.25}{\per\angstrom}$ along the equivalent ZU directions, arising from the falling part of B$_1$ and B$_2$ bulk bands (see Fig. \ref{fig:Fig3}b). The nice agreement with theoretical DFT bands is seen in panel 3h, where the crossing points of B$_1$ and B$_2$ bands with the isoenergy plane at 0.8 eV are shown. At this energy a six-fold star cannot be detected because bands along ZA stay at much higher energy (see Fig. \ref{fig:Fig3}b). Note that, while the tangential broadening of the experimental circles at $\SI{0.25}{\angstrom}^{-1}$ is due to rotational domains, the radial broadening is much higher than that of the parallel segments seen in Fig. \ref{fig:Fig3}e. This signals that this broadening is not purely experimental but reflects the collapse of the two crossing points of bulk B$_1$ and B$_2$. As expected from theory, the k-splitting of B$_1$ and B$_2$ bands at 0.8 eV BE is much smaller than that of S$_1$ and S$_2$ at the Fermi level, thus impeding to distinguish them because of the intrinsic experimental broadening of bands in our experiment. 

Finally, let us stress the link between the direction of the built-in FE polarization ($P_{up}$) and the observation of the k-splitting of S$_1$ and S$_2$ surface Rashba bands. In Fig. \ref{fig:Fig4}, we report surface bands calculated by DFT in case of a Te terminated GeTe surface without any relaxation. In this way it is possible to arbitrarily set the direction of the FE polarization, outward $\left(P_{up}\right)$ or inward $\left(P_{dn}\right)$, while in case of the relaxed surface the unique possibility is the more energetically stable $P_{up}$ \cite{Deringer2012}. Surface bands are identified plotting on the surface bandstructures, by means of red circles, the weight of the surface Te atoms orbital character. A huge Rashba splitting, compatible with experimental data, is seen only for the $P_{up}$ termination (Fig. \ref{fig:Fig4}a), thus underlying the connection between FE polarization and surface Rashba observed in GeTe. In fact, for termination with polarization $P_{dn}$ (Fig. \ref{fig:Fig4}b), surface bands are pushed towards the conduction bands, and clear Rashba-like surface features compatible with experimental data are missing. Note that the entity of this surface Rashba effect connected to S$_1$ and S$_2$ bands is very high. As shown in Fig. \ref{fig:Fig4}a, we estimate a maximum spin-splitting ($\Delta E$) of about 220 meV with a momentum offset ($\Delta k$) as large as $\SI{0.16}{\angstrom}$, for S$_1$ and S$_2$ bands along $k_y$ (ZU,$\overline{\Gamma} \overline{M}$). This value for the energy splitting is by far one of the largest ever reported so far for surfaces or interfaces \cite{LaShell1996,Nitta1997,Caviglia2010,Ishizaka2011,Crepaldi2012,Eremeev2012}. Note that, at variance with all other systems investigated so far, for GeTe the Rashba effect is linked to the ferroelectric polarization. This unique feature is highly appealing, as the spin chirality of Rashba bands could be exploited as a state variable electrically controlled and with the inherent stability associated to the ferroelectric remanence. 
\\
\indent To summarize, we have shown that GeTe band structure displays the characteristic features of a FERSC material. An intrinsic remanent outwards FE polarization has been detected by PFM, which provides the inversion symmetry breaking needed to observe Rashba effects. Surface Rashba bands linked to the outwards built-in FE polarization of $\alpha$-GeTe(111) films have been identified and clearly disentangled from bulk Rashba states. This work provides the first experimental investigation of the intriguing peculiarities of FERSC materials and paves the way to their use in novel architectures of spintronic devices.



\begin{acknowledgments}
We acknowledge M. Leone for his skillful technical assistance, M. Fenner for his assistance for PFM data, J. Krempasky, J. Minar and J. H. Dil for discussions. This work was partially funded by the Italian Ministry of Research through the project FIRB RBAP115AYN and by EU within the FP7 project PASTRY (GA 317746).
\end{acknowledgments}


\begin{thebibliography}{26}%
\makeatletter
\providecommand \@ifxundefined [1]{%
 \@ifx{#1\undefined}
}%
\providecommand \@ifnum [1]{%
 \ifnum #1\expandafter \@firstoftwo
 \else \expandafter \@secondoftwo
 \fi
}%
\providecommand \@ifx [1]{%
 \ifx #1\expandafter \@firstoftwo
 \else \expandafter \@secondoftwo
 \fi
}%
\providecommand \natexlab [1]{#1}%
\providecommand \enquote  [1]{``#1''}%
\providecommand \bibnamefont  [1]{#1}%
\providecommand \bibfnamefont [1]{#1}%
\providecommand \citenamefont [1]{#1}%
\providecommand \href@noop [0]{\@secondoftwo}%
\providecommand \href [0]{\begingroup \@sanitize@url \@href}%
\providecommand \@href[1]{\@@startlink{#1}\@@href}%
\providecommand \@@href[1]{\endgroup#1\@@endlink}%
\providecommand \@sanitize@url [0]{\catcode `\\12\catcode `\$12\catcode
  `\&12\catcode `\#12\catcode `\^12\catcode `\_12\catcode `\%12\relax}%
\providecommand \@@startlink[1]{}%
\providecommand \@@endlink[0]{}%
\providecommand \url  [0]{\begingroup\@sanitize@url \@url }%
\providecommand \@url [1]{\endgroup\@href {#1}{\urlprefix }}%
\providecommand \urlprefix  [0]{URL }%
\providecommand \Eprint [0]{\href }%
\providecommand \doibase [0]{http://dx.doi.org/}%
\providecommand \selectlanguage [0]{\@gobble}%
\providecommand \bibinfo  [0]{\@secondoftwo}%
\providecommand \bibfield  [0]{\@secondoftwo}%
\providecommand \translation [1]{[#1]}%
\providecommand \BibitemOpen [0]{}%
\providecommand \bibitemStop [0]{}%
\providecommand \bibitemNoStop [0]{.\EOS\space}%
\providecommand \EOS [0]{\spacefactor3000\relax}%
\providecommand \BibitemShut  [1]{\csname bibitem#1\endcsname}%
\let\auto@bib@innerbib\@empty
\bibitem [{\citenamefont {Manchon}(2014)}]{Manchon2014}%
  \BibitemOpen
  \bibfield  {author} {\bibinfo {author} {\bibfnamefont {A.}~\bibnamefont
  {Manchon}},\ }\href {http://dx.doi.org/10.1038/nphys2957} {\bibfield
  {journal} {\bibinfo  {journal} {Nat Phys}\ }\textbf {\bibinfo {volume}
  {10}},\ \bibinfo {pages} {340} (\bibinfo {year} {2014})}\BibitemShut
  {NoStop}%
\bibitem [{\citenamefont {Garello}\ \emph {et~al.}(2013)\citenamefont
  {Garello}, \citenamefont {Miron}, \citenamefont {Avci}, \citenamefont
  {Freimuth}, \citenamefont {Mokrousov}, \citenamefont {Blugel}, \citenamefont
  {Auffret}, \citenamefont {Boulle}, \citenamefont {Gaudin},\ and\
  \citenamefont {Gambardella}}]{Garello2013}%
  \BibitemOpen
  \bibfield  {author} {\bibinfo {author} {\bibfnamefont {K.}~\bibnamefont
  {Garello}}, \bibinfo {author} {\bibfnamefont {I.~M.}\ \bibnamefont {Miron}},
  \bibinfo {author} {\bibfnamefont {C.~O.}\ \bibnamefont {Avci}}, \bibinfo
  {author} {\bibfnamefont {F.}~\bibnamefont {Freimuth}}, \bibinfo {author}
  {\bibfnamefont {Y.}~\bibnamefont {Mokrousov}}, \bibinfo {author}
  {\bibfnamefont {S.}~\bibnamefont {Blugel}}, \bibinfo {author} {\bibfnamefont
  {S.}~\bibnamefont {Auffret}}, \bibinfo {author} {\bibfnamefont
  {O.}~\bibnamefont {Boulle}}, \bibinfo {author} {\bibfnamefont
  {G.}~\bibnamefont {Gaudin}}, \ and\ \bibinfo {author} {\bibfnamefont
  {P.}~\bibnamefont {Gambardella}},\ }\href
  {http://dx.doi.org/10.1038/nnano.2013.145} {\bibfield  {journal} {\bibinfo
  {journal} {Nat Nano}\ }\textbf {\bibinfo {volume} {8}},\ \bibinfo {pages}
  {587} (\bibinfo {year} {2013})}\BibitemShut {NoStop}%
\bibitem [{\citenamefont {Datta}\ and\ \citenamefont {Das}(1990)}]{Datta1990}%
  \BibitemOpen
  \bibfield  {author} {\bibinfo {author} {\bibfnamefont {S.}~\bibnamefont
  {Datta}}\ and\ \bibinfo {author} {\bibfnamefont {B.}~\bibnamefont {Das}},\
  }\href {\doibase http://dx.doi.org/10.1063/1.102730} {\bibfield  {journal}
  {\bibinfo  {journal} {Applied Physics Letters}\ }\textbf {\bibinfo {volume}
  {56}},\ \bibinfo {pages} {665} (\bibinfo {year} {1990})}\BibitemShut
  {NoStop}%
\bibitem [{\citenamefont {Koo}\ \emph {et~al.}(2009)\citenamefont {Koo},
  \citenamefont {Kwon}, \citenamefont {Eom}, \citenamefont {Chang},
  \citenamefont {Han},\ and\ \citenamefont {Johnson}}]{Koo2009}%
  \BibitemOpen
  \bibfield  {author} {\bibinfo {author} {\bibfnamefont {H.~C.}\ \bibnamefont
  {Koo}}, \bibinfo {author} {\bibfnamefont {J.~H.}\ \bibnamefont {Kwon}},
  \bibinfo {author} {\bibfnamefont {J.}~\bibnamefont {Eom}}, \bibinfo {author}
  {\bibfnamefont {J.}~\bibnamefont {Chang}}, \bibinfo {author} {\bibfnamefont
  {S.~H.}\ \bibnamefont {Han}}, \ and\ \bibinfo {author} {\bibfnamefont
  {M.}~\bibnamefont {Johnson}},\ }\href {\doibase 10.1126/science.1173667}
  {\bibfield  {journal} {\bibinfo  {journal} {Science}\ }\textbf {\bibinfo
  {volume} {325}},\ \bibinfo {pages} {1515} (\bibinfo {year} {2009})},\ \Eprint
  {http://arxiv.org/abs/http://www.sciencemag.org/content/325/5947/1515.full.pdf}
  {http://www.sciencemag.org/content/325/5947/1515.full.pdf} \BibitemShut
  {NoStop}%
\bibitem [{\citenamefont {Picozzi}(2014)}]{Picozzi2014}%
  \BibitemOpen
  \bibfield  {author} {\bibinfo {author} {\bibfnamefont {S.}~\bibnamefont
  {Picozzi}},\ }\href {\doibase 10.3389/fphy.2014.00010} {\bibfield  {journal}
  {\bibinfo  {journal} {Frontiers in Physics}\ }\textbf {\bibinfo {volume} {2}}
  (\bibinfo {year} {2014}),\ 10.3389/fphy.2014.00010}\BibitemShut {NoStop}%
\bibitem [{\citenamefont {Di~Sante}\ \emph {et~al.}(2013)\citenamefont
  {Di~Sante}, \citenamefont {Barone}, \citenamefont {Bertacco},\ and\
  \citenamefont {Picozzi}}]{DiSante2013}%
  \BibitemOpen
  \bibfield  {author} {\bibinfo {author} {\bibfnamefont {D.}~\bibnamefont
  {Di~Sante}}, \bibinfo {author} {\bibfnamefont {P.}~\bibnamefont {Barone}},
  \bibinfo {author} {\bibfnamefont {R.}~\bibnamefont {Bertacco}}, \ and\
  \bibinfo {author} {\bibfnamefont {S.}~\bibnamefont {Picozzi}},\ }\href
  {\doibase 10.1002/adma.201203199} {\bibfield  {journal} {\bibinfo  {journal}
  {Advanced Materials}\ }\textbf {\bibinfo {volume} {25}},\ \bibinfo {pages}
  {509} (\bibinfo {year} {2013})}\BibitemShut {NoStop}%
\bibitem [{\citenamefont {Chattopadhyay}\ \emph {et~al.}(1987)\citenamefont
  {Chattopadhyay}, \citenamefont {Boucherle},\ and\ \citenamefont
  {vonSchnering}}]{Chattopadhyay1987}%
  \BibitemOpen
  \bibfield  {author} {\bibinfo {author} {\bibfnamefont {T.}~\bibnamefont
  {Chattopadhyay}}, \bibinfo {author} {\bibfnamefont {J.~X.}\ \bibnamefont
  {Boucherle}}, \ and\ \bibinfo {author} {\bibfnamefont {H.~G.}\ \bibnamefont
  {vonSchnering}},\ }\href {http://stacks.iop.org/0022-3719/20/i=10/a=012}
  {\bibfield  {journal} {\bibinfo  {journal} {Journal of Physics C: Solid State
  Physics}\ }\textbf {\bibinfo {volume} {20}},\ \bibinfo {pages} {1431}
  (\bibinfo {year} {1987})}\BibitemShut {NoStop}%
\bibitem [{\citenamefont {Rabe}\ and\ \citenamefont
  {Joannopoulos}(1987)}]{Rabe1987}%
  \BibitemOpen
  \bibfield  {author} {\bibinfo {author} {\bibfnamefont {K.~M.}\ \bibnamefont
  {Rabe}}\ and\ \bibinfo {author} {\bibfnamefont {J.~D.}\ \bibnamefont
  {Joannopoulos}},\ }\href {\doibase 10.1103/PhysRevB.36.6631} {\bibfield
  {journal} {\bibinfo  {journal} {Phys. Rev. B}\ }\textbf {\bibinfo {volume}
  {36}},\ \bibinfo {pages} {6631} (\bibinfo {year} {1987})}\BibitemShut
  {NoStop}%
\bibitem [{\citenamefont {Shaltaf}\ \emph {et~al.}(2008)\citenamefont
  {Shaltaf}, \citenamefont {Durgun}, \citenamefont {Raty}, \citenamefont
  {Ghosez},\ and\ \citenamefont {Gonze}}]{Shaltaf2008}%
  \BibitemOpen
  \bibfield  {author} {\bibinfo {author} {\bibfnamefont {R.}~\bibnamefont
  {Shaltaf}}, \bibinfo {author} {\bibfnamefont {E.}~\bibnamefont {Durgun}},
  \bibinfo {author} {\bibfnamefont {J.-Y.}\ \bibnamefont {Raty}}, \bibinfo
  {author} {\bibfnamefont {P.}~\bibnamefont {Ghosez}}, \ and\ \bibinfo {author}
  {\bibfnamefont {X.}~\bibnamefont {Gonze}},\ }\href {\doibase
  10.1103/PhysRevB.78.205203} {\bibfield  {journal} {\bibinfo  {journal} {Phys.
  Rev. B}\ }\textbf {\bibinfo {volume} {78}},\ \bibinfo {pages} {205203}
  (\bibinfo {year} {2008})}\BibitemShut {NoStop}%
\bibitem [{\citenamefont {Binasch}\ \emph {et~al.}(1989)\citenamefont
  {Binasch}, \citenamefont {Gr\"unberg}, \citenamefont {Saurenbach},\ and\
  \citenamefont {Zinn}}]{Binasch1989}%
  \BibitemOpen
  \bibfield  {author} {\bibinfo {author} {\bibfnamefont {G.}~\bibnamefont
  {Binasch}}, \bibinfo {author} {\bibfnamefont {P.}~\bibnamefont {Gr\"unberg}},
  \bibinfo {author} {\bibfnamefont {F.}~\bibnamefont {Saurenbach}}, \ and\
  \bibinfo {author} {\bibfnamefont {W.}~\bibnamefont {Zinn}},\ }\href {\doibase
  10.1103/PhysRevB.39.4828} {\bibfield  {journal} {\bibinfo  {journal} {Phys.
  Rev. B}\ }\textbf {\bibinfo {volume} {39}},\ \bibinfo {pages} {4828}
  (\bibinfo {year} {1989})}\BibitemShut {NoStop}%
\bibitem [{\citenamefont {Giussani}\ \emph {et~al.}(2012)\citenamefont
  {Giussani}, \citenamefont {Perumal}, \citenamefont {Hanke}, \citenamefont
  {Rodenbach}, \citenamefont {Riechert},\ and\ \citenamefont
  {Calarco}}]{Giussani2012}%
  \BibitemOpen
  \bibfield  {author} {\bibinfo {author} {\bibfnamefont {A.}~\bibnamefont
  {Giussani}}, \bibinfo {author} {\bibfnamefont {K.}~\bibnamefont {Perumal}},
  \bibinfo {author} {\bibfnamefont {M.}~\bibnamefont {Hanke}}, \bibinfo
  {author} {\bibfnamefont {P.}~\bibnamefont {Rodenbach}}, \bibinfo {author}
  {\bibfnamefont {H.}~\bibnamefont {Riechert}}, \ and\ \bibinfo {author}
  {\bibfnamefont {R.}~\bibnamefont {Calarco}},\ }\href {\doibase
  10.1002/pssb.201200367} {\bibfield  {journal} {\bibinfo  {journal} {physica
  status solidi (b)}\ }\textbf {\bibinfo {volume} {249}},\ \bibinfo {pages}
  {1939} (\bibinfo {year} {2012})}\BibitemShut {NoStop}%
\bibitem [{\citenamefont {Pawley}\ \emph {et~al.}(1966)\citenamefont {Pawley},
  \citenamefont {Cochran}, \citenamefont {Cowley},\ and\ \citenamefont
  {Dolling}}]{Pawley1966}%
  \BibitemOpen
  \bibfield  {author} {\bibinfo {author} {\bibfnamefont {G.~S.}\ \bibnamefont
  {Pawley}}, \bibinfo {author} {\bibfnamefont {W.}~\bibnamefont {Cochran}},
  \bibinfo {author} {\bibfnamefont {R.~A.}\ \bibnamefont {Cowley}}, \ and\
  \bibinfo {author} {\bibfnamefont {G.}~\bibnamefont {Dolling}},\ }\href
  {\doibase 10.1103/PhysRevLett.17.753} {\bibfield  {journal} {\bibinfo
  {journal} {Phys. Rev. Lett.}\ }\textbf {\bibinfo {volume} {17}},\ \bibinfo
  {pages} {753} (\bibinfo {year} {1966})}\BibitemShut {NoStop}%
\bibitem [{\citenamefont {Matsunaga}\ \emph {et~al.}(2008)\citenamefont
  {Matsunaga}, \citenamefont {Morita}, \citenamefont {Kojima}, \citenamefont
  {Yamada}, \citenamefont {Kifune}, \citenamefont {Kubota}, \citenamefont
  {Tabata}, \citenamefont {Kim}, \citenamefont {Kobata}, \citenamefont
  {Ikenaga},\ and\ \citenamefont {Kobayashi}}]{Matsunaga2008}%
  \BibitemOpen
  \bibfield  {author} {\bibinfo {author} {\bibfnamefont {T.}~\bibnamefont
  {Matsunaga}}, \bibinfo {author} {\bibfnamefont {H.}~\bibnamefont {Morita}},
  \bibinfo {author} {\bibfnamefont {R.}~\bibnamefont {Kojima}}, \bibinfo
  {author} {\bibfnamefont {N.}~\bibnamefont {Yamada}}, \bibinfo {author}
  {\bibfnamefont {K.}~\bibnamefont {Kifune}}, \bibinfo {author} {\bibfnamefont
  {Y.}~\bibnamefont {Kubota}}, \bibinfo {author} {\bibfnamefont
  {Y.}~\bibnamefont {Tabata}}, \bibinfo {author} {\bibfnamefont {J.-J.}\
  \bibnamefont {Kim}}, \bibinfo {author} {\bibfnamefont {M.}~\bibnamefont
  {Kobata}}, \bibinfo {author} {\bibfnamefont {E.}~\bibnamefont {Ikenaga}}, \
  and\ \bibinfo {author} {\bibfnamefont {K.}~\bibnamefont {Kobayashi}},\ }\href
  {\doibase http://dx.doi.org/10.1063/1.2901187} {\bibfield  {journal}
  {\bibinfo  {journal} {Journal of Applied Physics}\ }\textbf {\bibinfo
  {volume} {103}},\ \bibinfo {eid} {093511} (\bibinfo {year}
  {2008})}\BibitemShut {NoStop}%
\bibitem [{\citenamefont {Kadlec}\ \emph {et~al.}(2011)\citenamefont {Kadlec},
  \citenamefont {Kadlec}, \citenamefont {Ku\ifmmode~\check{z}\else
  \v{z}\fi{}el},\ and\ \citenamefont {Petzelt}}]{Kadlec2011}%
  \BibitemOpen
  \bibfield  {author} {\bibinfo {author} {\bibfnamefont {F.}~\bibnamefont
  {Kadlec}}, \bibinfo {author} {\bibfnamefont {C.}~\bibnamefont {Kadlec}},
  \bibinfo {author} {\bibfnamefont {P.}~\bibnamefont {Ku\ifmmode~\check{z}\else
  \v{z}\fi{}el}}, \ and\ \bibinfo {author} {\bibfnamefont {J.}~\bibnamefont
  {Petzelt}},\ }\href {\doibase 10.1103/PhysRevB.84.205209} {\bibfield
  {journal} {\bibinfo  {journal} {Phys. Rev. B}\ }\textbf {\bibinfo {volume}
  {84}},\ \bibinfo {pages} {205209} (\bibinfo {year} {2011})}\BibitemShut
  {NoStop}%
\bibitem [{\citenamefont {Gervacio-Arciniega}\ \emph
  {et~al.}(2012)\citenamefont {Gervacio-Arciniega}, \citenamefont {Prokhorov},
  \citenamefont {Espinoza-Beltrán},\ and\ \citenamefont
  {Trapaga}}]{Gervacio-Arciniega2012}%
  \BibitemOpen
  \bibfield  {author} {\bibinfo {author} {\bibfnamefont {J.~J.}\ \bibnamefont
  {Gervacio-Arciniega}}, \bibinfo {author} {\bibfnamefont {E.}~\bibnamefont
  {Prokhorov}}, \bibinfo {author} {\bibfnamefont {F.~J.}\ \bibnamefont
  {Espinoza-Beltrán}}, \ and\ \bibinfo {author} {\bibfnamefont
  {G.}~\bibnamefont {Trapaga}},\ }\href {\doibase
  http://dx.doi.org/10.1063/1.4746087} {\bibfield  {journal} {\bibinfo
  {journal} {Journal of Applied Physics}\ }\textbf {\bibinfo {volume} {112}},\
  \bibinfo {eid} {052018} (\bibinfo {year} {2012})}\BibitemShut {NoStop}%
\bibitem [{\citenamefont {Polking}\ \emph {et~al.}(2012)\citenamefont
  {Polking}, \citenamefont {Han}, \citenamefont {Yourdkhani}, \citenamefont
  {Petkov}, \citenamefont {Kisielowski}, \citenamefont {Volkov}, \citenamefont
  {Zhu}, \citenamefont {Caruntu}, \citenamefont {Paul~Alivisatos},\ and\
  \citenamefont {Ramesh}}]{Polking2012}%
  \BibitemOpen
  \bibfield  {author} {\bibinfo {author} {\bibfnamefont {M.~J.}\ \bibnamefont
  {Polking}}, \bibinfo {author} {\bibfnamefont {M.-G.}\ \bibnamefont {Han}},
  \bibinfo {author} {\bibfnamefont {A.}~\bibnamefont {Yourdkhani}}, \bibinfo
  {author} {\bibfnamefont {V.}~\bibnamefont {Petkov}}, \bibinfo {author}
  {\bibfnamefont {C.~F.}\ \bibnamefont {Kisielowski}}, \bibinfo {author}
  {\bibfnamefont {V.~V.}\ \bibnamefont {Volkov}}, \bibinfo {author}
  {\bibfnamefont {Y.}~\bibnamefont {Zhu}}, \bibinfo {author} {\bibfnamefont
  {G.}~\bibnamefont {Caruntu}}, \bibinfo {author} {\bibfnamefont
  {A.}~\bibnamefont {Paul~Alivisatos}}, \ and\ \bibinfo {author} {\bibfnamefont
  {R.}~\bibnamefont {Ramesh}},\ }\href {http://dx.doi.org/10.1038/nmat3371}
  {\bibfield  {journal} {\bibinfo  {journal} {Nat Mater}\ }\textbf {\bibinfo
  {volume} {11}},\ \bibinfo {pages} {700} (\bibinfo {year} {2012})}\BibitemShut
  {NoStop}%
\bibitem [{\citenamefont {Kolobov}\ \emph {et~al.}(2014)\citenamefont
  {Kolobov}, \citenamefont {Kim}, \citenamefont {Giussani}, \citenamefont
  {Fons}, \citenamefont {Tominaga}, \citenamefont {Calarco},\ and\
  \citenamefont {Gruverman}}]{Kolobov2014}%
  \BibitemOpen
  \bibfield  {author} {\bibinfo {author} {\bibfnamefont {A.~V.}\ \bibnamefont
  {Kolobov}}, \bibinfo {author} {\bibfnamefont {D.~J.}\ \bibnamefont {Kim}},
  \bibinfo {author} {\bibfnamefont {A.}~\bibnamefont {Giussani}}, \bibinfo
  {author} {\bibfnamefont {P.}~\bibnamefont {Fons}}, \bibinfo {author}
  {\bibfnamefont {J.}~\bibnamefont {Tominaga}}, \bibinfo {author}
  {\bibfnamefont {R.}~\bibnamefont {Calarco}}, \ and\ \bibinfo {author}
  {\bibfnamefont {A.}~\bibnamefont {Gruverman}},\ }\href {\doibase
  http://dx.doi.org/10.1063/1.4881735} {\bibfield  {journal} {\bibinfo
  {journal} {APL Materials}\ }\textbf {\bibinfo {volume} {2}},\  (\bibinfo
  {year} {2014})}\BibitemShut {NoStop}%
\bibitem [{\citenamefont {Deringer}\ \emph {et~al.}(2012)\citenamefont
  {Deringer}, \citenamefont {Lumeij},\ and\ \citenamefont
  {Dronskowski}}]{Deringer2012}%
  \BibitemOpen
  \bibfield  {author} {\bibinfo {author} {\bibfnamefont {V.~L.}\ \bibnamefont
  {Deringer}}, \bibinfo {author} {\bibfnamefont {M.}~\bibnamefont {Lumeij}}, \
  and\ \bibinfo {author} {\bibfnamefont {R.}~\bibnamefont {Dronskowski}},\
  }\bibfield  {booktitle} {\emph {\bibinfo {booktitle} {The Journal of Physical
  Chemistry C}},\ }\href {\doibase 10.1021/jp304455z} {\bibfield  {journal}
  {\bibinfo  {journal} {J. Phys. Chem. C}\ }\textbf {\bibinfo {volume} {116}},\
  \bibinfo {pages} {15801} (\bibinfo {year} {2012})}\BibitemShut {NoStop}%
\bibitem [{\citenamefont {Wang}\ \emph {et~al.}(2014)\citenamefont {Wang},
  \citenamefont {Boschker}, \citenamefont {Bruyer}, \citenamefont {Di~Sante},
  \citenamefont {Picozzi}, \citenamefont {Perumal}, \citenamefont {Giussani},
  \citenamefont {Riechert},\ and\ \citenamefont {Calarco}}]{Wang0}%
  \BibitemOpen
  \bibfield  {author} {\bibinfo {author} {\bibfnamefont {R.}~\bibnamefont
  {Wang}}, \bibinfo {author} {\bibfnamefont {J.~E.}\ \bibnamefont {Boschker}},
  \bibinfo {author} {\bibfnamefont {E.}~\bibnamefont {Bruyer}}, \bibinfo
  {author} {\bibfnamefont {D.}~\bibnamefont {Di~Sante}}, \bibinfo {author}
  {\bibfnamefont {S.}~\bibnamefont {Picozzi}}, \bibinfo {author} {\bibfnamefont
  {K.}~\bibnamefont {Perumal}}, \bibinfo {author} {\bibfnamefont
  {A.}~\bibnamefont {Giussani}}, \bibinfo {author} {\bibfnamefont
  {H.}~\bibnamefont {Riechert}}, \ and\ \bibinfo {author} {\bibfnamefont
  {R.}~\bibnamefont {Calarco}},\ }\href {\doibase 10.1021/jp507183f} {\bibfield
   {journal} {\bibinfo  {journal} {The Journal of Physical Chemistry C}\ }
  (\bibinfo {year} {2014}),\ 10.1021/jp507183f},\ \Eprint
  {http://arxiv.org/abs/http://dx.doi.org/10.1021/jp507183f}
  {http://dx.doi.org/10.1021/jp507183f} \BibitemShut {NoStop}%
\bibitem [{\citenamefont {Wuttig}\ \emph {et~al.}(2007)\citenamefont {Wuttig},
  \citenamefont {Lusebrink}, \citenamefont {Wamwangi}, \citenamefont {Welnic},
  \citenamefont {Gilleszen},\ and\ \citenamefont {Dronskowski}}]{Wuttig2007}%
  \BibitemOpen
  \bibfield  {author} {\bibinfo {author} {\bibfnamefont {M.}~\bibnamefont
  {Wuttig}}, \bibinfo {author} {\bibfnamefont {D.}~\bibnamefont {Lusebrink}},
  \bibinfo {author} {\bibfnamefont {D.}~\bibnamefont {Wamwangi}}, \bibinfo
  {author} {\bibfnamefont {W.}~\bibnamefont {Welnic}}, \bibinfo {author}
  {\bibfnamefont {M.}~\bibnamefont {Gilleszen}}, \ and\ \bibinfo {author}
  {\bibfnamefont {R.}~\bibnamefont {Dronskowski}},\ }\href
  {http://dx.doi.org/10.1038/nmat1807} {\bibfield  {journal} {\bibinfo
  {journal} {Nat Mater}\ }\textbf {\bibinfo {volume} {6}},\ \bibinfo {pages}
  {122} (\bibinfo {year} {2007})}\BibitemShut {NoStop}%
\bibitem [{\citenamefont {LaShell}\ \emph {et~al.}(1996)\citenamefont
  {LaShell}, \citenamefont {McDougall},\ and\ \citenamefont
  {Jensen}}]{LaShell1996}%
  \BibitemOpen
  \bibfield  {author} {\bibinfo {author} {\bibfnamefont {S.}~\bibnamefont
  {LaShell}}, \bibinfo {author} {\bibfnamefont {B.~A.}\ \bibnamefont
  {McDougall}}, \ and\ \bibinfo {author} {\bibfnamefont {E.}~\bibnamefont
  {Jensen}},\ }\href {\doibase 10.1103/PhysRevLett.77.3419} {\bibfield
  {journal} {\bibinfo  {journal} {Phys. Rev. Lett.}\ }\textbf {\bibinfo
  {volume} {77}},\ \bibinfo {pages} {3419} (\bibinfo {year}
  {1996})}\BibitemShut {NoStop}%
\bibitem [{\citenamefont {Nitta}\ \emph {et~al.}(1997)\citenamefont {Nitta},
  \citenamefont {Akazaki}, \citenamefont {Takayanagi},\ and\ \citenamefont
  {Enoki}}]{Nitta1997}%
  \BibitemOpen
  \bibfield  {author} {\bibinfo {author} {\bibfnamefont {J.}~\bibnamefont
  {Nitta}}, \bibinfo {author} {\bibfnamefont {T.}~\bibnamefont {Akazaki}},
  \bibinfo {author} {\bibfnamefont {H.}~\bibnamefont {Takayanagi}}, \ and\
  \bibinfo {author} {\bibfnamefont {T.}~\bibnamefont {Enoki}},\ }\href
  {http://link.aps.org/doi/10.1103/PhysRevLett.78.1335} {\bibfield  {journal}
  {\bibinfo  {journal} {Phys. Rev. Lett.}\ }\textbf {\bibinfo {volume} {78}},\
  \bibinfo {pages} {1335} (\bibinfo {year} {1997})}\BibitemShut {NoStop}%
\bibitem [{\citenamefont {Caviglia}\ \emph {et~al.}(2010)\citenamefont
  {Caviglia}, \citenamefont {Gabay}, \citenamefont {Gariglio}, \citenamefont
  {Reyren}, \citenamefont {Cancellieri},\ and\ \citenamefont
  {Triscone}}]{Caviglia2010}%
  \BibitemOpen
  \bibfield  {author} {\bibinfo {author} {\bibfnamefont {A.~D.}\ \bibnamefont
  {Caviglia}}, \bibinfo {author} {\bibfnamefont {M.}~\bibnamefont {Gabay}},
  \bibinfo {author} {\bibfnamefont {S.}~\bibnamefont {Gariglio}}, \bibinfo
  {author} {\bibfnamefont {N.}~\bibnamefont {Reyren}}, \bibinfo {author}
  {\bibfnamefont {C.}~\bibnamefont {Cancellieri}}, \ and\ \bibinfo {author}
  {\bibfnamefont {J.-M.}\ \bibnamefont {Triscone}},\ }\href
  {http://link.aps.org/doi/10.1103/PhysRevLett.104.126803} {\bibfield
  {journal} {\bibinfo  {journal} {Phys. Rev. Lett.}\ }\textbf {\bibinfo
  {volume} {104}},\ \bibinfo {pages} {126803} (\bibinfo {year}
  {2010})}\BibitemShut {NoStop}%
\bibitem [{\citenamefont {Ishizaka}\ \emph {et~al.}(2011)\citenamefont
  {Ishizaka}, \citenamefont {Bahramy}, \citenamefont {Murakawa}, \citenamefont
  {Sakano}, \citenamefont {Shimojima}, \citenamefont {Sonobe}, \citenamefont
  {Koizumi}, \citenamefont {Shin}, \citenamefont {Miyahara}, \citenamefont
  {Kimura}, \citenamefont {Miyamoto}, \citenamefont {Okuda}, \citenamefont
  {Namatame}, \citenamefont {Taniguchi}, \citenamefont {Arita}, \citenamefont
  {Nagaosa}, \citenamefont {Kobayashi}, \citenamefont {Murakami}, \citenamefont
  {Kumai}, \citenamefont {Kaneko}, \citenamefont {Onose},\ and\ \citenamefont
  {Tokura}}]{Ishizaka2011}%
  \BibitemOpen
  \bibfield  {author} {\bibinfo {author} {\bibfnamefont {K.}~\bibnamefont
  {Ishizaka}}, \bibinfo {author} {\bibfnamefont {M.~S.}\ \bibnamefont
  {Bahramy}}, \bibinfo {author} {\bibfnamefont {H.}~\bibnamefont {Murakawa}},
  \bibinfo {author} {\bibfnamefont {M.}~\bibnamefont {Sakano}}, \bibinfo
  {author} {\bibfnamefont {T.}~\bibnamefont {Shimojima}}, \bibinfo {author}
  {\bibfnamefont {T.}~\bibnamefont {Sonobe}}, \bibinfo {author} {\bibfnamefont
  {K.}~\bibnamefont {Koizumi}}, \bibinfo {author} {\bibfnamefont
  {S.}~\bibnamefont {Shin}}, \bibinfo {author} {\bibfnamefont {H.}~\bibnamefont
  {Miyahara}}, \bibinfo {author} {\bibfnamefont {A.}~\bibnamefont {Kimura}},
  \bibinfo {author} {\bibfnamefont {K.}~\bibnamefont {Miyamoto}}, \bibinfo
  {author} {\bibfnamefont {T.}~\bibnamefont {Okuda}}, \bibinfo {author}
  {\bibfnamefont {H.}~\bibnamefont {Namatame}}, \bibinfo {author}
  {\bibfnamefont {M.}~\bibnamefont {Taniguchi}}, \bibinfo {author}
  {\bibfnamefont {R.}~\bibnamefont {Arita}}, \bibinfo {author} {\bibfnamefont
  {N.}~\bibnamefont {Nagaosa}}, \bibinfo {author} {\bibfnamefont
  {K.}~\bibnamefont {Kobayashi}}, \bibinfo {author} {\bibfnamefont
  {Y.}~\bibnamefont {Murakami}}, \bibinfo {author} {\bibfnamefont
  {R.}~\bibnamefont {Kumai}}, \bibinfo {author} {\bibfnamefont
  {Y.}~\bibnamefont {Kaneko}}, \bibinfo {author} {\bibfnamefont
  {Y.}~\bibnamefont {Onose}}, \ and\ \bibinfo {author} {\bibfnamefont
  {Y.}~\bibnamefont {Tokura}},\ }\href {http://dx.doi.org/10.1038/nmat3051}
  {\bibfield  {journal} {\bibinfo  {journal} {Nat Mater}\ }\textbf {\bibinfo
  {volume} {10}},\ \bibinfo {pages} {521} (\bibinfo {year} {2011})}\BibitemShut
  {NoStop}%
\bibitem [{\citenamefont {Crepaldi}\ \emph {et~al.}(2012)\citenamefont
  {Crepaldi}, \citenamefont {Moreschini}, \citenamefont {Autès}, \citenamefont
  {Tournier-Colletta}, \citenamefont {Moser}, \citenamefont {Virk},
  \citenamefont {Berger}, \citenamefont {Bugnon}, \citenamefont {Chang},
  \citenamefont {Kern}, \citenamefont {Bostwick}, \citenamefont {Rotenberg},
  \citenamefont {Yazyev},\ and\ \citenamefont {Grioni}}]{Crepaldi2012}%
  \BibitemOpen
  \bibfield  {author} {\bibinfo {author} {\bibfnamefont {A.}~\bibnamefont
  {Crepaldi}}, \bibinfo {author} {\bibfnamefont {L.}~\bibnamefont
  {Moreschini}}, \bibinfo {author} {\bibfnamefont {G.}~\bibnamefont {Autès}},
  \bibinfo {author} {\bibfnamefont {C.}~\bibnamefont {Tournier-Colletta}},
  \bibinfo {author} {\bibfnamefont {S.}~\bibnamefont {Moser}}, \bibinfo
  {author} {\bibfnamefont {N.}~\bibnamefont {Virk}}, \bibinfo {author}
  {\bibfnamefont {H.}~\bibnamefont {Berger}}, \bibinfo {author} {\bibfnamefont
  {P.}~\bibnamefont {Bugnon}}, \bibinfo {author} {\bibfnamefont {Y.~J.}\
  \bibnamefont {Chang}}, \bibinfo {author} {\bibfnamefont {K.}~\bibnamefont
  {Kern}}, \bibinfo {author} {\bibfnamefont {A.}~\bibnamefont {Bostwick}},
  \bibinfo {author} {\bibfnamefont {E.}~\bibnamefont {Rotenberg}}, \bibinfo
  {author} {\bibfnamefont {O.~V.}\ \bibnamefont {Yazyev}}, \ and\ \bibinfo
  {author} {\bibfnamefont {M.}~\bibnamefont {Grioni}},\ }\href
  {http://link.aps.org/doi/10.1103/PhysRevLett.109.096803} {\bibfield
  {journal} {\bibinfo  {journal} {Phys. Rev. Lett.}\ }\textbf {\bibinfo
  {volume} {109}},\ \bibinfo {pages} {096803} (\bibinfo {year}
  {2012})}\BibitemShut {NoStop}%
\bibitem [{\citenamefont {Eremeev}\ \emph {et~al.}(2012)\citenamefont
  {Eremeev}, \citenamefont {Nechaev}, \citenamefont {Koroteev}, \citenamefont
  {Echenique},\ and\ \citenamefont {Chulkov}}]{Eremeev2012}%
  \BibitemOpen
  \bibfield  {author} {\bibinfo {author} {\bibfnamefont {S.~V.}\ \bibnamefont
  {Eremeev}}, \bibinfo {author} {\bibfnamefont {I.~A.}\ \bibnamefont
  {Nechaev}}, \bibinfo {author} {\bibfnamefont {Y.~M.}\ \bibnamefont
  {Koroteev}}, \bibinfo {author} {\bibfnamefont {P.~M.}\ \bibnamefont
  {Echenique}}, \ and\ \bibinfo {author} {\bibfnamefont {E.~V.}\ \bibnamefont
  {Chulkov}},\ }\href {http://link.aps.org/doi/10.1103/PhysRevLett.108.246802}
  {\bibfield  {journal} {\bibinfo  {journal} {Phys. Rev. Lett.}\ }\textbf
  {\bibinfo {volume} {108}},\ \bibinfo {pages} {246802} (\bibinfo {year}
  {2012})}\BibitemShut {NoStop}%
\end{thebibliography}
%

\end{document}